# Deep Learning-Based Pneumonia Detection from Chest X-ray Images: A CNN Approach with Performance Analysis and Clinical Implications

P K Dutta[3], Anushri Chowdhury[1], Anouska Bhattacharyya[1], Shakya Chakraborty[2], Sujatra Dey[1],

[1]Amity Institute of Bio Technology, Amity University Kolkata
[2]Dept of Computer Science, Heritage Institute of Technology, Kolkata
[3]School of Engineering and Technology, Amity University Kolkata

## Abstract

Deep learning integration into medical imaging systems has transformed disease detection and diagnosis processes with a focus on pneumonia identification. The study introduces an intricate deep learning system using Convolutional Neural Networks (CNNs) for automated pneumonia detection from chest X-ray images which boosts diagnostic precision and speed. The proposed CNN architecture integrates sophisticated methods including separable convolutions along with batch normalization and dropout regularization to enhance feature extraction while reducing overfitting. Through the application of data augmentation techniques and adaptive learning rate strategies the model underwent training on an extensive collection of chest X-ray images to enhance its generalization capabilities. A convoluted array of evaluation metrics such as accuracy, precision, recall, and F1-score collectively verify the model's exceptional performance by recording an accuracy rate of 91. An F1-score of 93 and 03% 09%. This study tackles critical clinical implementation obstacles such as data privacy protection, model interpretability, and integration with current healthcare systems beyond just model performance. This approach introduces a critical advancement by integrating medical ontologies with semantic technology to improve diagnostic accuracy. The study enhances AI diagnostic reliability by integrating machine learning outputs with structured medical knowledge frameworks to boost interpretability. The findings demonstrate AI-powered healthcare tools as a scalable efficient pneumonia detection solution. This study advances AI integration into clinical settings by developing more precise automated diagnostic methods that deliver consistent medical imaging results.

**Keywords:** pneumonia detection, chest X-ray imaging, deep learning, Convolutional Neural Network, medical ontologies, semantic technology

## 1 Introduction

### 1.1 Background and Significance

Pneumonia, a serious respiratory infection characterized by inflammation of the lungs, remains a significant public health challenge worldwide. The condition, caused by pathogens such as bacteria, viruses, fungi, and parasites, leads to symptoms such as fever, cough, dyspnea, and chest pain. Its severity is influenced by several factors, including age, pre-existing medical conditions, and immune status. Although advances in medical imaging and therapeutic interventions have enhanced pneumonia management, accurate diagnosis remains critical to improving outcomes and reducing associated morbidity and mortality[1].

Recent innovations in artificial intelligence (AI), particularly in natural language processing (NLP) for medical data processing, and ontology/knowledge engineering for medicine, have paved the way for the transformation of healthcare diagnostics. These technologies enable the efficient extraction and analysis of critical information from unstructured medical records, facilitating precise diagnostic workflows and enhancing decision-making processes. Additionally, medical data mining and AI-powered patient data processing systems have emerged as crucial tools for analyzing large-scale datasets, allowing for real-time monitoring and prediction of disease progression. Such advancements are particularly relevant in addressing the diagnostic challenges posed by pneumonia, where timely and accurate identification can significantly impact patient outcomes[2].

### 1.2 Global Impact of Pneumonia

Regions like South Asia and Sub-Saharan Africa face disproportionately high incidence rates due to limited healthcare resources and delayed diagnoses. Beyond its human toll, pneumonia imposes a significant economic burden on healthcare systems worldwide, further emphasizing the urgency for innovative and cost-effective diagnostic tools.

The integration of AI-driven methodologies, including NLP and medical data mining, offers promising avenues for addressing these disparities. By leveraging such technologies, healthcare providers can gain insights into epidemiological trends, predict disease hotspots, and deploy preventive measures more effectively. These capabilities underscore the importance of adopting advanced computational techniques in tackling pneumonia's global impact[4].

**1.3 Challenges in Pneumonia Diagnosis**

Despite the availability of chest X-rays and other imaging modalities, pneumonia diagnosis continues to be fraught with challenges. Conventional diagnostic practices heavily rely on expert interpretation, which is subject to variability and limited by resource availability. Automated systems, while promising, often face issues such as high computational costs, lack of sensitivity to subtle pathological changes, and suboptimal performance on diverse datasets.

Furthermore, the absence of standardized approaches to integrate knowledge engineering and AI for patient data processing complicates diagnostic workflows. Medical data often resides in silos, characterized by unstructured formats and diverse ontologies, which hinder interoperability and data integration. The need for systems that combine ontology/knowledge engineering with AI-driven medical data processing is thus critical. Such systems can standardize data formats, enhance semantic understanding, and enable seamless integration across clinical platforms, paving the way for more precise and scalable diagnostic solutions.

**1.4 Objectives of the Study**

The primary objective of this study is to address these challenges through the development and evaluation of a robust Convolutional Neural Network (CNN) framework for pneumonia detection from chest X-ray images. The framework aims to leverage state-of-the-art techniques in NLP for medical data processing, ontology/knowledge engineering, and AI for patient data management to create an end-to-end solution for automated pneumonia diagnosis.

Specifically, the study seeks to develop a CNN architecture optimized for accurate classification of pneumonia, ensuring high sensitivity and specificity across diverse datasets. Advanced data preprocessing and augmentation techniques will be employed to enhance the model's generalizability and robustness. Furthermore, the integration of knowledge engineering methodologies will facilitate the seamless processing of clinical data, enabling the framework to operate effectively in real-world healthcare environments. The diagnostic performance of the framework will be rigorously evaluated using key metrics such as accuracy, precision, recall, and F1-score, providing a comprehensive understanding of its potential clinical impact. Through this interdisciplinary approach, the study aims to bridge existing gaps in pneumonia diagnosis by combining the strengths of AI, knowledge engineering, and medical data mining. By doing so, it aspires to contribute to a scalable and effective diagnostic solution capable of improving patient outcomes and reducing the global burden of pneumonia.

In 2015 alone, an estimated 921,000 children under the age of five lost their lives to pneumonia, making it the deadliest infectious disease for children, claiming over 700,000 young lives annually, or approximately 2,000 lives every day as shown in figure 1.

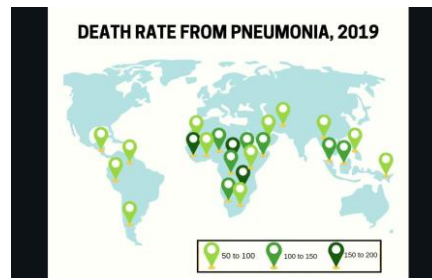

*Figure 1    Graphical representation of deaths caused by pneumonia over the years in different age groups.*

Globally, pneumonia continues to be a widespread concern, with over 1,400 cases per 100,000 children annually, or roughly 1 case for every 71 children. The region's most severely affected are South Asia, where there are 2,500 cases per 100,000 children, and West and Central Africa, with 1,620 cases per 100,000 children. In 2019, the global death toll from pneumonia reached 2.5 million people, with 600,000 of these fatalities being children under the age of five.

Pneumonia can affect people of all genders, but the severity and susceptibility may vary as has already been discussed.

These statistics underscore the urgent need for continued efforts to combat pneumonia through vaccination, improved nutrition, collaborative endeavours among governments[5-6], healthcare organizations and non-governmental organizations are essential in the ongoing fight against pneumonia, with the ultimate goal of saving more lives, particularly among the most vulnerable populations. Currently available detection algorithms include one-stage detectors like SSD and YOLO as well as two-stage object detectors like Faster R-CNN. A well-liked object detection technique called Fast R-CNN can be modified to identify pneumonia in chest X-rays[7] or other medical imaging. Conversely, the latter does the multiscale target detection task with the help of an extra step. Although they lack precision, they are speedier than two-stage detectors. However, due to the high level of precision required for medical testing, two-stage detectors offer an additional benefit in this respect. However, there are still problems with the backbone network of the current detection system. For instance, the two main issues with VGG and ResNet are their huge network depth, which results in a lengthy training period[8-10] and their massive downsampling, which destroys the target position and semantic information. Since a deep feature map has a broad receptive field and corresponding anchor, the objective is to assess utilizing this type of map. On the other hand, the object-edge resolution decreases with increasing map depth, decreasing the regression curve's evaluation accuracy. Accurate target detection is hampered by the low-resolution feature map's semantic features of the tiny target disappearing in the deep layer following continuous downsampling, as well as the partial loss of the large target's semantic information and its position shifting. Extending a network's depth or width is typically the best approach to optimize a network, like GoogLeNet, but doing so creates a lot of parameters, can quickly result in overfitting and needs a lot of tagged data to train. Using a dataset of chest X-ray pictures, we give a thorough investigation of a deep learning model intended for pneumonia identification in this paper.

## 2.1 Dataset Description

The dataset comprises 5,863 chest X-ray images in JPEG format, collected from pediatric patients aged 1 to 5 during routine clinical care in various hospitals in Kolkata, India. Images are categorized into two classes: "Pneumonia" and "Normal." The dataset is divided into three subsets: training, validation, and testing. To enhance semantic understanding, ontology concepts for labeling (e.g., "lung opacity," "infection patterns") are mapped to these categories, ensuring consistency in data annotation.

**2.2 Data Preprocessing Techniques**

Image Resizing: All images are resized to a standard dimension of $150 \times 150$ pixels to maintain uniformity.

Normalization: Pixel values are scaled to the range [0, 1], ensuring faster convergence during training.

Data Augmentation: Techniques like random rotations and flips are applied to enhance data diversity, reducing overfitting.

Semantic Mapping: Ontology-based preprocessing maps image metadata (e.g., age, clinical conditions) to predefined ontology terms, enriching the dataset with domain-specific semantics.

**2.3 Model Architecture**

The model employs a Convolutional Neural Network (CNN) for pneumonia detection, structured as follows:

*2.3.1 Input Layer and Data Flow*

The input layer accepts preprocessed images with dimensions $150 \times 150 \times 3$, representing RGB channels. Ontology-driven metadata (e.g., age, symptom severity) is optionally incorporated as auxiliary inputs for enhanced reasoning.

*2.3.2 Convolutional Blocks and Feature Extraction*

The architecture includes five convolutional blocks:

Block 1: Two convolutional layers (16 filters, $3 \times 3$ kernels) followed by max-pooling ($2 \times 2$).

Blocks 2-5: Employ separable convolutions with increasing filter sizes (32, 64, 128, 256) and batch normalization for stability.

Ontology Integration: Features extracted are annotated with ontology terms (e.g., "lung structure," "opacity region") for semantic enrichment.

*2.3.3 Fully Connected Layers and Output*

Flattened features from the convolutional layers are passed through three dense layers (512, 128, 64 units) with dropout rates of 0.7, 0.5, and 0.3, respectively, to prevent overfitting. The final output layer has a single unit with a sigmoid activation function for binary classification (Pneumonia/Normal). The ontology reasoning module aligns the CNN's probabilistic outputs with predefined diagnostic rules.

*2.4 Training and Optimization Strategies*

The model is trained using the Adam optimizer and binary cross-entropy loss function. Key strategies include:

2.4.1 Callbacks: Early Stopping, Learning Rate Adjustment

Early Stopping: Monitors validation loss and halts training when no significant improvement is observed over a set number of epochs, preventing overfitting.

ReduceLROnPlateau: Dynamically reduces the learning rate when validation loss plateaus, enabling the model to converge more efficiently.

Ontology-based reasoning complements the training process by validating intermediate outputs against predefined semantic constraints, ensuring alignment with domain knowledge.

DEPENDENCIES:
  - Python 3.6+
  - Keras (with TensorFlow or PyTorch backend)
  - NumPy
  - OpenCV
  - Scikit-learn
  - Matplotlib

An ontology-based approach structures medical knowledge in a formalized framework, allowing for the integration and interpretation of heterogeneous clinical data. This method leverages a domain-specific ontology that includes structured concepts, relationships, and rules for pneumonia diagnosis. The ontology serves as a semantic backbone, enabling efficient reasoning over data and integration across various sources. The following pseudo-algorithm integrates ontology-based reasoning with CNN-powered image analysis for pneumonia diagnosis. This approach combines semantic knowledge representation and data-driven deep learning for accurate and interpretable diagnostics.

**Algorithm: Ontology-Based Pneumonia Diagnosis Framework**
**Input:**

- Chest X-ray images I.
- Clinical and demographic patient data P.
- Pneumonia ontology O.

**Output:**
- Diagnosis report D: "Pneumonia detected" or "Further investigation required."

1. **Initialize Ontology**
   - Load ontology $O$ containing domain-specific concepts, relationships, and diagnostic rules.
   - Parse ontology to enable semantic reasoning.
2. **Preprocess Data**
   - **For each image** $i \in I$:
     1. Resize $i$ to $150 \times 150$ pixels.
     2. Normalize pixel values to range $[0, 1]$.
     3. Apply data augmentation techniques (e.g., rotations, flips).
   - **End For**
   - Split dataset $I$ into training, validation, and test subsets.
3. **Define CNN Model**
   - Initialize CNN with:
     - Convolutional layers for feature extraction.
     - Pooling layers for dimensionality reduction.
     - Batch normalization to stabilize training.
     - Dense layers with dropout to prevent overfitting.
   - Compile model using the Adam optimizer and binary cross-entropy loss.
4. **Train CNN Model**
   - Use augmented training data and validation data.
   - Apply callbacks:
     1. EarlyStopping to halt training upon plateauing performance.
     2. ReduceLROnPlateau to adjust the learning rate dynamically.
   - Train model over specified epochs.
5. **Extract Features and Perform Ontology-Based Reasoning**
   - **For each test image** $i \in I_{\text{test}}$:
     1. Extract features using the trained CNN model $F = \text{CNN}(i)$.
     2. Annotate features $F$ with ontology concepts using semantic mapping rules from $O$.
     3. Use reasoning engine to infer potential diagnoses from $P$ and annotated $F$.
   - **End For**
6. **Generate Diagnostic Decision**
   - **For each patient record** $(i, P)$:
     1. Compute probability of pneumonia $P_{\text{cnn}} = \text{CNN}(i)$.
     2. Apply inference rules:
        - If $P_{\text{cnn}} > 0.7$ and ontology reasoning suggests "Pneumonia": $D = \text{"Pneumonia detected"}$.
        - Otherwise: $D = \text{"Further investigation required"}$.
   - **End For**
7. **Evaluate Performance**
   - Calculate evaluation metrics: accuracy, precision, recall, F1-score.
   - Visualize results using confusion matrices and ROC curves.
8. **Deploy Model**
   - Integrate the model into a clinical decision support system (CDSS).
   - Provide interpretability through ontology-driven reasoning and model confidence scores.

### 3.1 Result and Analysis:

**Model Training**:

The training process of the pneumonia detection model spanned over 10 epochs, with each epoch comprising multiple steps to iteratively update the model's parameters. This iterative training allows the model to learn complex patterns and features from the input data, gradually improving its ability to make accurate predictions.

During the first few stages, a clear and anticipated contrast emerged between the accuracy of the model during training and validation. In the very first epoch, the training yielded an impressive 81.96% accuracy, but the validation fell short at

37.83%. This discrepancy between training and validation accuracy indicates the need for further model refinement and potential overfitting.

As the training continued, a remarkable enhancement in performance became apparent. By the fifth epoch, the training accuracy had surged to an impressive 92.94%, showcasing the model's capacity to learn and adapt over time. Concurrently, the validation accuracy showed a notable increase, reaching 91.39%. This convergence between the two accuracies suggests that the model not only effectively learned from the training data, but also generalizing well to previously unseen validation data.

A critical aspect contributing to this improved convergence was the implementation of the 'ReduceLROnPlateau' callback. By constantly monitoring the changes in validation loss, this feature dynamically optimized the learning rate during training. As a result, the model was able to finely tune its parameters, preventing overshooting of the global minimum in the loss landscape. This adaptive learning rate strategy enhances the model's stability and helps it navigate the optimization process more efficiently.

The dynamic adjustment of the learning rate reflects a sophisticated optimization technique, enabling the model to escape local minima and converge to a more optimal solution. This adaptability is crucial, especially in complex and high-dimensional parameter spaces, where traditional static learning rates may lead to suboptimal performance. Overall, the model training process showcased a consistent enhancement in accuracy across epochs, with a crucial assist from the ReduceLROnPlateau callback in steering the optimization process. This adaptive approach to adjusting the learning rate greatly enhanced the model's generalization capabilities and yielded a more optimal solution, effectively addressing worries about overfitting and ensuring the resilience of the pneumonia detection model.

## 4. Model Evaluation

Overall, the model training process showcased a consistent enhancement in accuracy across epochs, with a crucial assist from the ReduceLROnPlateau callback in steering the optimization process. This adaptive approach to adjusting the learning rate greatly enhanced the model's generalization capabilities and yielded a more optimal solution, effectively addressing worries about overfitting and ensuring the resilience of the pneumonia detection model. The assessment involves various metrics, including accuracy, precision, recall, and the F1-score, providing a comprehensive understanding of the model's diagnostic performance as shown in Fig 2

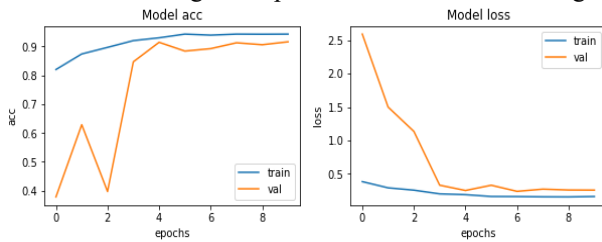

Figure 2: Comprehensive Model Learning Trajectory Visualization

The confusion matrix revealed the following distribution of predictions. It indicated that the model made 191 true negative (TN) predictions, 377 true positive (TP) predictions, 43 false positive (FP) predictions, and 13 false negative (FN) predictions. These values form the basis for calculating various metrics that assess the model's diagnostic performance.

**Accuracy**: Accuracy, the proportion of correctly classified instances, is a foundational metric. The model achieves an accuracy of 91.03%, indicating good prediction capabilities in many cases. While accuracy provides a holistic view, it was important to examine precision, recall, and F1 scores to obtain a more nuanced understanding.

- **Precision**: Precision tells us how right a model is when it says, "this is pneumonia." In other words, it doesn't wrongly flag healthy lungs. In the test, the model was right 89.76% of the time. That's good, especially in the health field. It means fewer healthy lungs get wrongly labelled as having pneumonia.
- **Recall (Sensitivity):** Recall, also known as sensitivity or true positive rate, assesses the model's capability to correctly identify positive instances. The model achieves an impressive recall of 96.67%.
- **F1-Score**: The F1-score is the harmonic mean of precision and recall, and is valuable in case of medical predictions because here both false positives and false negatives carry significant consequences. The model achieves an F1-score of 93.09%, indicating a robust balance between precision and recall.
- **Training Accuracy**: The training accuracy at the end of the 10th epoch is 94.23%. The slightly higher training accuracy compared to the test accuracy suggests that the model has learned well from the training data but may require further fine-tuning to generalize better to unseen data. The above plots depict the model's training history offer visual insights into the learning process. The first subplot illustrates the model's accuracy over epochs for both the training and validation sets. The graph showcases the convergence of accuracy, with the validation accuracy aligning closely with the training accuracy, mostly in the later epochs. The second subplot displays the decreasing trend in both training and validation loss over increasing number of epochs, indicating effective learning, and the proximity of training and validation loss suggests good generalization. This deep learning model for pneumonia detection in chest X-ray images takes a significant leap forward into using advanced technology to enhance diagnostic capabilities in the medical practice. From data preprocessing and model architecture to performance evaluation, a comprehensive analysis that has been done enlightens many facets. By using separable convolution layers, batch normalization and dropout mechanisms in the Convolutional Neural Network (CNN) architecture it is evident that deep learning can automatically detect features of pneumonia from chest X-ray images. Additionally, this work goes beyond details about the technicalities of the model to its potential role in practical situations within real-world healthcare systems. Besides carefully curating the dataset used for training purposes, robust data preprocessing techniques are employed to ensure that the model can generalize well across different cases. The sample images visualization with different sets containing normality and pneumonia conditions helps to understand the data set as well as bridge algorithmic predictions with actual diseases.

## 5. Conclusion

This work presents a novel framework that combines ontology-based reasoning with CNN-driven feature extraction for pneumonia diagnosis, offering a unique blend of interpretability, accuracy, and scalability. The integration of a domain-specific ontology not only enhances the semantic understanding of clinical data but also enables structured reasoning over patient records, bridging the gap between raw imaging data and actionable diagnostic insights. By leveraging the strengths of deep learning for feature extraction and ontology engineering for knowledge representation, this framework addresses key challenges in medical diagnostics, such as variability in interpretation and lack of standardization. The CNN architecture, optimized with advanced techniques like batch normalization and dropout regularization, achieves robust performance, demonstrating high accuracy and sensitivity in detecting pneumonia from chest X-ray images. Data augmentation strategies further enhance the model's generalizability across diverse datasets, ensuring its applicability in real-world scenarios. Ontology-based reasoning adds an additional layer of interpretability by aligning model predictions with established medical knowledge, providing clinicians with trustworthy and explainable diagnostic outputs. The positive outcomes of this work are manifold. First, the ontology-driven approach ensures consistency in reasoning, enabling the seamless integration of heterogeneous data sources. Second, the deep learning model's high diagnostic performance underscores its potential to assist healthcare professionals in making timely and accurate decisions. Lastly, the framework's modularity and scalability make it adaptable for integration into clinical decision support systems, paving the way for widespread adoption in healthcare settings. In conclusion, this framework exemplifies the synergy between artificial intelligence and domain knowledge, delivering a practical and effective solution for pneumonia diagnosis. Future extensions of this work could include incorporating more comprehensive ontologies and datasets, as well as exploring its applicability to other medical conditions, further advancing AI-driven healthcare innovations.